\documentclass[aps,pre,twocolumn]{revtex4-1}
\usepackage{amsmath,bm,epsfig}
\usepackage{epsfig, natbib}
\usepackage{bm}
\newcommand{\B}[1]{{\bm{#1}}}

\newcommand{\pa}{\partial}
\usepackage[latin1]{inputenc}
\usepackage[dvips]{color}

\begin{document}
\title{An intrinsic nonlinear scale governs oscillations in rapid fracture}
\author{Tamar Goldman$^1$, Roi Harpaz$^2$, Eran Bouchbinder$^2$ and Jay Fineberg$^1$}
\affiliation{$^1$The Racah Institute of Physics, The Hebrew University of Jerusalem, Jerusalem 91904, Israel\\
$^2$Chemical Physics Department, Weizmann Institute of Science, Rehovot 76100, Israel}

\date{\today}
\begin{abstract}
{When branching is suppressed, rapid cracks undergo a dynamic instability from a straight to an oscillatory path at a critical velocity $v_c$. In a systematic experimental study using a wide range of different brittle materials, we first show how the opening profiles of straight cracks scale with the size $\ell_{nl}$ of the nonlinear zone surrounding a crack's tip.  We then show, for all materials tested, that $v_c$ is both a fixed fraction of the shear speed and, moreover, that the instability wavelength is proportional to $\ell_{nl}$. These findings directly verify recent theoretical predictions and suggest that the nonlinear zone is not passive, but rather is closely linked to rapid crack instabilities.}
\end{abstract}
\maketitle

Since their discovery, a fundamental understanding of the origin of rapid crack instabilities \cite{Ravi-Chandar.84b,Fineberg.91,Deegan.02.rubber,Livne.07,Scheibert.2010} has proven to be very elusive. The dynamics of single straight cracks are well described \cite{Sharon.99,Goldman.2010, BFM10} by Linear Elastic Fracture Mechanics (LEFM) \cite{Freund.90,Marder.91.prl}.
This theoretical framework predicts singular crack tip fields and describes a crack's dynamics as a balance between the elastic energy flux into the tip region and the energy dissipated at the tip. LEFM, however, cannot explain rapid crack instabilities and accompanying non-trivial crack patterns, without additional assumptions or physical insights about the near-tip region where linear elasticity breaks down.

There have been a number of notable attempts to describe crack instabilities by supplementing or extending LEFM in various ways. These include phase-field models \cite{Aranson.00,Karma2004,Henry2004,Henry.08,Spatschek.06, SBK11}, cohesive-zone models \cite{Falk.01,Miller.99,Langer.98}, models based on the ``Principle of Local Symmetry" \cite{Adda-Bedia.99, Movchan.05, Bouchbinder.07}, energy conservation bounds on crack branching \cite{Eshelby.70, Adda-Bedia.05} and models based on non-linear constitutive behavior near the crack tip \cite{Buehler2006, Bouchbinder.09b}. Although many of these models are qualitatively consistent with both experimental and numerical observations, decisive quantitative experiments that are able to differentiate between them are lacking.

Here we focus on the oscillatory instability in rapid brittle fracture in which a straight crack becomes unstable to sinusoidal path oscillations \cite{Livne.07}. The onset of these oscillations was observed at a critical velocity $v_c$ of about 90\% of the shear wave speed $c_s$, when the micro-branching instability \cite{Ravi-Chandar.84b,Fineberg.91} was suppressed. This instability is particularly intriguing since it involves a finite instability wavelength at onset that is independent of either system geometry or loading conditions. This suggests the existence of an intrinsic scale that can not exist in linear elastic solutions for cracks, which are scale-free.

Recently, a theory describing this instability was proposed \cite{Bouchbinder.09b}. This theory is based on the existence of a dynamic non-linear lengthscale $\ell_{nl}$, where linear elasticity breaks down and material nonlinearities become significant due to the large deformation near a crack's tip \cite{Livne.08,Livne.2010,Bouchbinder.08a,Bouchbinder.09a}. The basic idea behind this approach is that in the presence of a finite $\ell_{nl}$, causality implies that the singular LEFM fields lag behind the actual tip location with a delay of $\tau_d \propto \ell_{nl}$. This led to a high-velocity oscillatory instability with the following properties: (i) the scaled critical velocity for the onset of oscillations $v_c/c_s$ is material independent (ii) the oscillation wavelength, $\lambda_{osc}$, is proportional to $\ell_{nl}$.

In this Letter we investigate the rapid fracture of a variety of different brittle gels, whose mechanical properties vary over a wide range. We first demonstrate that the opening profiles of straight cracks collapse onto a single velocity-dependent form, when scaled by the size of the nonlinear elastic zone, as predicted by \cite{Bouchbinder.08a}. We then show that the oscillatory instability is triggered in each material at the {\em same} scaled value of $v_c/c_s$ and, moreover, that the instability wavelength indeed scales with $\ell_{nl}$, confirming the theoretical predictions of \cite{Bouchbinder.09b}.

Our experiments were performed using polyacrylamide gels which are transparent, homogeneous, brittle, incompressible elastomers. The dynamics of rapid cracks in these neo-Hookean materials are identical to those observed in other brittle amorphous materials (e.g. glass, PMMA). Due to the low elastic moduli of these soft materials, the wave speeds and corresponding crack velocities are nearly 3 orders of magnitude \cite{Livne.05} lower than in conventional materials. This enables us to slow down the fracture process while obtaining detailed measurements of rapid cracks at unprecedented scaled velocities.

We control the gels' physical properties by varying their chemical composition \cite{Tanaka.03}. We varied the total monomer concentration (by weight) between $14.2\%\!-\!32.4\%$, cross-linker concentration between $2.7\%\!-\!4.6\%$  and polymer initiators in the range $0.03\%\!-\!0.06\%$. In what follows we will label each gel by its shear modulus, $\mu$ ($33\!<\!\mu\!<\!187$\,kPa) and fracture energy at the critical velocity, $\Gamma(v_c)$ ($24\!<\!\Gamma(v_c)\!<\!60$\,J/m$^2$). $\Gamma$ is defined as the amount of energy dissipated per unit crack extension and sample thickness. The details of these gel compositions are provided in \cite{supplementary}. Typical dimensions of the gels used were ($x\times y\times z$) ($130\times 130\times 0.2$)mm and ($200\times 200\times 0.2$)mm, where $x$, $y$ and $z$ are, respectively, the propagation, loading and thickness directions.

Experiments were performed as in \cite{Livne.07} by imposing uniaxial tensile loading via constant displacement in the vertical ($y$) direction. Once a desired strain $\epsilon$ was reached, a guillotine was used to initiate fracture at the sample's edge, midway between the vertical boundaries. To negate boundary effects on the crack tip prior to the onset of the oscillatory instability, the applied strain levels were selected such that the crack could reach very high velocities ($\sim 0.9c_s$) before traversing half of the sample's size. For experimentally feasible system sizes, this entails strains $\epsilon$ in the range $6\!-\!18\%$. The crack tip opening displacement (CTOD) of the moving crack was measured with a high speed camera focussed on an ($x\times y$) area of $60\times 9.5\!-\!19$mm with $1280\times 200\!-\!400$ pixel resolution. Successive photographs were taken at between $2490/15000$ frames/s with a $2\mu$s exposure time. Multiple exposures were utilized, when needed. The micro-branching instability was suppressed (as in \cite{Livne.07}) by setting the gel thickness to $160\!-\!220\mu$m. In all experiments analyzed, no micro-branches occurred in the regions of interest. Post-fracture $xy$ profiles were measured via an optical scanner with 300dpi resolution.

Let us now consider a simple straight crack moving at velocity $v$ under constant tensile loading, prior to any instability. According to LEFM, the CTOD has a parabolic shape whose curvature $a(v)$ is inversely proportional to the instantaneous value of $\Gamma(v)$ \cite{Freund.90}. This characteristic parabolic form is indeed experimentally measured at points that are at a distance not too close to the crack tip. Sufficiently near the crack tip, regions of very high strain are encountered. The resulting nonlinear elastic effects shift the actual crack tip by a distance $\delta$ from the apex of the parabolic form defined by LEFM \cite{Livne.08}.
The dissipative zone adjacent to the tip is also contained within $\delta$. In gels, the dissipative zone is significantly smaller than the size of the non-linear elastic deformation zone \cite{Livne.2010}.

The strain levels imposed in our measurements suggest that the CTOD predicted by LEFM should be calculated with respect to the background strain $\epsilon$. To this end, we consider the energy functional describing our incompressible gels under plane stress conditions \cite{Knowles.83}
\begin{equation}
\label{NH} U(\B F)= \frac{\mu}{2}\left[F_{ij}F_{ij}+\det(\B
F)^{-2}-3\right] \ ,
\end{equation}
where $\B F\!=\!\nabla\B \varphi$ is the deformation gradient and $\B x'\!=\!\B \varphi(\B x)$ is a mapping between a reference (undeformed) configuration $\B x$ and a deformed one $\B x'$. For our uniaxial loading we have $\varphi_x\!=\!(1+\epsilon)^{-1/2}x+u_x$ and $\varphi_y\!=\!(1+\epsilon)y+u_y$, where $\B u$ is the displacement field due to the presence of a crack. Using the stress measure $\B s\!=\!\pa_{\B F} U(\B F)$, the momentum balance equation reads $\nabla \cdot \B s \!=\!\rho \pa_{tt}{\B \varphi}$, where $\rho$ is the mass density. The traction-free boundary conditions on the crack faces take the form $s_{xy}(r,\theta\!=\!\pm\pi)\!=\!s_{yy}(r,\theta\!=\!\pm\pi)\!=\!0$, where $(r,\theta)$ is a polar coordinate system moving with the crack tip and $\theta\!=\!0$ is the propagation direction. Linearizing these equations with respect to $\B u$ and solving the resulting equations numerically near the tip of a crack moving at a steady velocity $v$, the solution takes the form \cite{supplementary}

\begin{eqnarray}
\hspace{-0.5cm}&&u_x(r,\theta; v,\epsilon)=\frac{K_I(v,\epsilon) \sqrt{r}}{4\mu\sqrt{2\pi}}\Omega_x(\theta;v,\epsilon)\!+\!\frac{T_x(\epsilon)\,r\cos\theta}{\mu},\nonumber\\
\label{firstO}
\hspace{-0.5cm}&&u_y(r,\theta; v,\epsilon)=\frac{K_I(v,\epsilon)\sqrt{r}}{4\mu\sqrt{2\pi}}\Omega_y(\theta;v,\epsilon)\!+\!\frac{T_y(\epsilon)\,r\sin\theta}{\mu} \ ,
\end{eqnarray}
where $K_I$ is the stress intensity factor, $\B T$ is a traction vector known as the T-stress and $\B \Omega$ is a universal angular function \cite{supplementary}. In the limit $\epsilon \!\rightarrow\! 0$ the standard LEFM solution is recovered \cite{Freund.90}.
Equation (\ref{firstO}) can be used to relate the parabolic crack tip curvature $a(v,\epsilon)$ and the
fracture energy $\Gamma(v)$, yielding
\begin{equation}
\label{gamma_SoL}
\Gamma(v)=\mu \left[(1+\epsilon)^{-1/2}+ \frac{T_x(\epsilon)}{\mu} \right]\dfrac{A(v;\epsilon)}{\Omega_y^2(\pi;v,\epsilon)}\dfrac{1}{a(v,\epsilon)} \ ,
\end{equation}
where $T_x(\epsilon)$ and $A(v;\epsilon)$ are given in \cite{supplementary}.

%%%%%%% FIGURE 1 %%%%%%%%%%%%%%%%%%
\begin{figure}
\centering \epsfig{width=.49\textwidth ,file=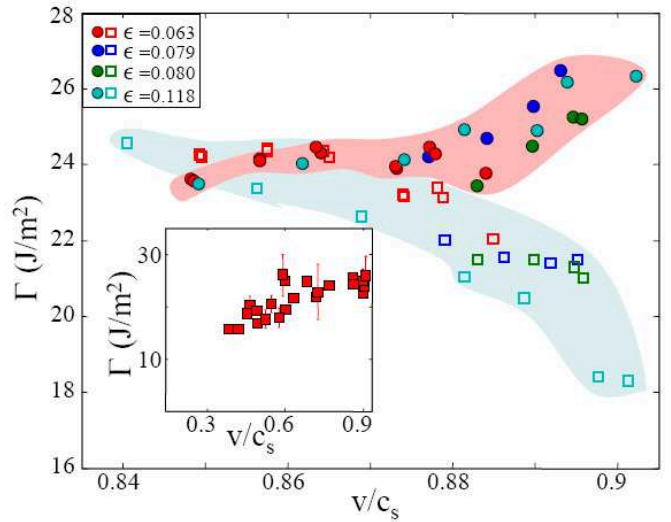}
\caption{(Color online) Comparison of the fracture energy, $\Gamma(v)$, calculated from (squares) CTOD measurements using the LEFM solution \cite{Freund.90} and (circles) using the extension of LEFM for finite strain given by Eq. (\ref{gamma_SoL}). The gel used has $\mu=110$kPa. (inset) Measurements of $\Gamma (v)$ obtained by velocity profiles in a strip geometry with a more compliant ($\mu=36$kPa) gel \cite{Goldman.2010}.}
\label{fig-gamma}
\end{figure}
%%%%%%%%%%%%%%%%%%%%%%%%%%%%%%%%%%%

The deviations of Eq. (\ref{gamma_SoL}) from the LEFM predictions are very small for low strains ($\epsilon\!<\!6\%$), justifying their omission in previous studies \cite{Livne.08,Livne.2010,Bouchbinder.08a,Bouchbinder.09a}. However, as $\epsilon$ is increased, significant corrections to the LEFM predictions appear for high crack velocities. In Fig. \ref{fig-gamma} we compare values of $\Gamma(v)$ derived from measured CTOD's using both Eq. (\ref{gamma_SoL}) and the analogous LEFM relation. While, at low velocities, the differences are insignificant, for $v>0.85c_s$ the $\Gamma(v)$ curves diverge significantly. The LEFM relation (blue-shading in Fig. \ref{fig-gamma}) yields both a large systematic variation of $\Gamma(v)$ for different $\epsilon$ values and a substantial decrease of $\Gamma$ with $v$. The former contradicts the expectation that $\Gamma(v)$ is a material-dependent function whose value should not reflect the background strain. Moreover, the sharp decrease in $\Gamma(v)$ at high velocities is surprising. One would expect a nearly constant value of $\Gamma(v)$ in this narrow range of velocities, as obtained in independent measurements of $\Gamma(v)$ (using an ``infinite strip" geometry \cite{Goldman.2010}) for a similar material (Fig. \ref{fig-gamma} - inset). The use of Eq. (\ref{gamma_SoL}) both eliminates the apparent dependence of $\Gamma$ with $\epsilon$ and indeed reveals the slowly increasing $\Gamma$ with $v$, consistent with the direct measurements presented in the inset. We were able to significantly decrease the  $2\!-\!4\%$ experimental uncertainty in $v/c_s$, by varying $c_s$ (within experimental uncertainty) to minimize the variance of the mean value of $\Gamma(v)$ over the range $0.85\!<\!v/c_s\!<\!0.9$. $\Gamma(v)$, obtained by this procedure, is presented within the red-shaded data in Fig. \ref{fig-gamma}. The collapse of the data together with the resulting slow increase of $\Gamma(v)$ with $v$ as expected from \cite{Goldman.2010}, justifies this procedure, which is used to determine $\Gamma(v)$ in what follows.

\begin{figure}
\centering \epsfig{width=.49\textwidth ,file=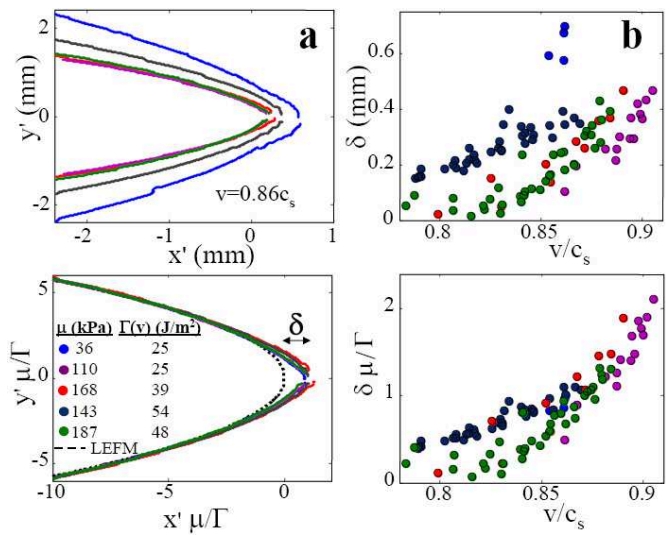}
\caption{(Color online)(a) Top: Measurements of the CTOD for 5 different materials (legend below) at $\epsilon\!\simeq\! 0.08$ and $v\!=\!0.86c_s$. $(x',y')$ are the coordinates in the laboratory (deformed) frame, to be distinguished from the reference (undeformed) frame $(x,y)$. Bottom: When scaled by $\Gamma(v)/\mu$, these curves collapse to a single function. Far from the tip at $(x',y')=(0,0)$ the CTOD is parabolic (dashed line), but strongly deviates from this form at a scale $\delta$, defined as the distance from the apex of these parabola to the crack tip. (b) Top: $\delta(v/c_s)$ for the 5 different materials in (a). Bottom: approximate collapse of the $\delta(v/c_s)$ curves when scaled by $\Gamma(v)/\mu$
.}\label{fig-scaling}
\end{figure}

For each $v/c_s$, Fig. \ref{fig-scaling}a demonstrates that scaling lengths by $\Gamma/\mu$ collapses the CTOD's of different materials to a single function. LEFM predicts that this should occur for the parabolic CTOD's away from the crack tip. It is {\em not} obvious, however, that data collapse should occur in the near-tip region defined by $\delta$, as this is a wholly independent regime. Data collapse with $\Gamma/\mu$ in the weakly nonlinear regime (i.e. cubic expansion of $U$ in the metric strain measure $\B E \!=\! \case{1}{2}(\B F^T \B F \!-\!\B I)$ \cite{supplementary}) was predicted for neo-Hookean materials \cite{Bouchbinder.08a,Livne.2010}, where second order elastic coefficients are order $\mu$. (In analogous scaling for other materials these coefficients may significantly differ from $\mu$ \cite{supplementary}.) A perfect data collapse would indicate that this is the {\em only} significant scale in the system. High-resolution measurements of $\delta(v)$, presented in Fig. \ref{fig-scaling}b for 5 different materials, provide a stringent test of this scaling. While the widely spread raw data (Fig. \ref{fig-scaling}b-top) indeed undergo an approximate collapse when scaled by $\Gamma/\mu$, the imperfect collapse for small values of scaled $\delta$ indicates that an additional, much smaller, scale exists. We surmise that this additional scale could be related to either the strongly nonlinear elastic region or the dissipation zone \cite{Livne.2010}.

%%%%%%%% FIGURE 2 %%%%%%%%%%%%%%%%%%
\begin{figure}
\centering \epsfig{width=.49\textwidth ,file=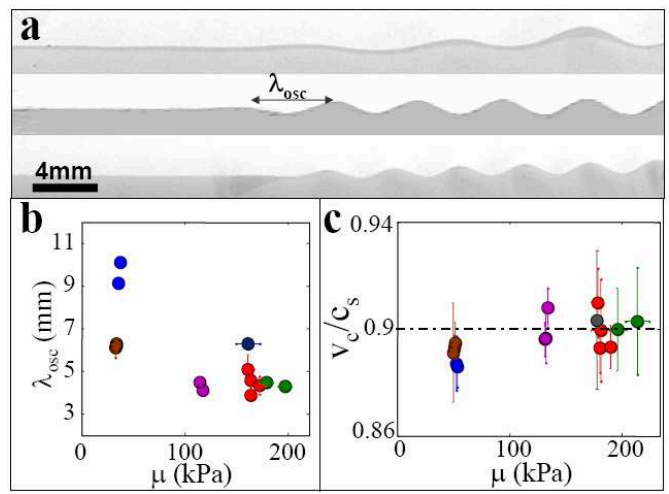}
\caption{(Color online) (a) Typical photographs of the $xy$ profiles of fracture surfaces at the onset of the oscillatory instability; from top to bottom: $\mu=36$kPa, $143$kPa, $168$kPa.
(b) The oscillation wavelength, $\lambda_{osc}$ changes significantly with the material, as characterized by $\mu$. (c) The scaled critical velocity, $v_c\cong0.9c_s$, is constant. $v_c$ is defined as the maximal velocity prior to the instability onset in each material. Symbol colors correspond to the legend of Fig. \ref{fig-scaling}.}
\label{osc_intro}
\end{figure}

%%%%%%%%%%%%%%%%%%%%%%%%%%%%%%%%%%%%
We now turn to the oscillatory instability. As shown in Fig. \ref{osc_intro}, the wavelength of the first oscillation, $\lambda_{osc}$, is strongly material-dependent, varying by over a factor of $2.5$ in different materials (Fig. \ref{osc_intro}b). In each material there is a well-defined velocity $v_c$ for the onset of the instability. As predicted by \cite{Bouchbinder.09b}, Fig. \ref{osc_intro}c shows that $v_c$, when scaled by $c_s$, has the nearly constant value of $v_c\!=\!0.9c_s$, in each of the 6 materials studied.
%%%%%%%% FIGURE 3 %%%%%%%%%%%%%%%%%%
%%%%%%%%%%%%%%%%%%%%%%%%%%%%%%%%%%%%%

\begin{figure}
\centering \epsfig{width=.49\textwidth ,file=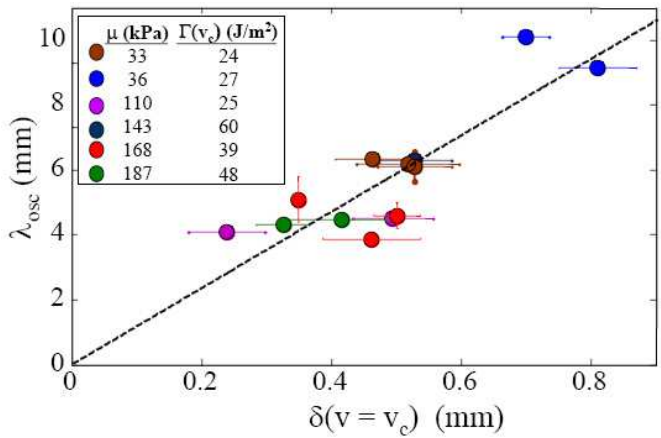}
\caption{(Color online) Comparison between the non-linear length-scale
$\delta(v\!=\!v_c)$ and the oscillation wavelength $\lambda_{osc}$. Note that the different combinations of $\mu$, $\Gamma$, and $\epsilon$ are used to produce $\sim 15$ independent measurements. The dashed line is a guide to the eye.}
\label{nl-scales}
\end{figure}

What is the origin of the instability wavelength? Fig. \ref{osc_intro} confirms that $\lambda_{osc}$ is not related to details of the experimental system. In experiments with identical conditions, $\lambda_{osc}$ varied widely with the material used. In \cite{Bouchbinder.09b}, $\lambda_{osc}$ was predicted to be proportional to the size of the nonlinear zone $\ell_{nl}$. Here we use $\delta(v\!=\!v_c)$ to estimate $\ell_{nl}$ at the critical velocity $v_c$ for different materials. The obvious advantage of doing this is that $\delta(v)$ is {\em directly measurable}, and hence the theoretical prediction of \cite{Bouchbinder.09b} can be recast as a relation between two directly measurable quantities, $\lambda_{osc}$ and $\delta(v\!=\!v_c)$. In Fig. \ref{nl-scales} we plot $\lambda_{osc}$ vs. $\delta(v\!=\!v_c)$ for the 6 materials used. We indeed find that $\delta$ is directly proportional to $\lambda_{osc}$, as predicted in \cite{Bouchbinder.09b}. Moreover, the constant of proportionality between $\lambda_{osc}$ and $\delta(v\!=\!v_c)$ in Fig. \ref{nl-scales} is consistent with the analysis of \cite{Bouchbinder.09b, supplementary}.

We note that the weakly nonlinear estimate of $\ell_{nl}\!\propto\!\Gamma/\mu$ \cite{Bouchbinder.08a} is also linearly related to $\lambda_{osc}$. In contrast to Fig. \ref{nl-scales}, however, this linear relation involves an offset corresponding to $\sim 100-300\mu$m. This scale is also apparent in the imperfect data collapse in Fig. \ref{fig-scaling}b, suggesting that $\delta$ includes length-scales such as the strongly nonlinear contributions to the nonlinear elastic zone and/or the scale of the ``dissipative zone'' at the crack tip which are beyond the perturbative estimate of Eq. (\ref{NH}) used in \cite{Bouchbinder.08a}.

%%%%%%%%%%%%%%%%%%%%%%%%%%%%%%%%%%%%
%%%%%%%%%%%%%%%%%%%%%%%%%%%%%%%%%%%%

In conclusion, our results conclusively demonstrate that the oscillatory instability of fast brittle cracks indeed involves an intrinsic scale that is governed, in a large part, by the nonlinear elastic zone surrounding the crack tip. The size of this zone quantitatively agrees with the predictions of \cite{Bouchbinder.09b}. These results indicate that the nonlinear (and dissipative) zones surrounding the tip of a moving crack are not ``passive'' objects that are simply ``dragged along'' by the crack tip. Instead, as suggested by \cite{Buehler2006, Livne.07, Bouchbinder.09b}, this region may play an active role in destabilizing crack motion. The demonstration of this presented in this work is, therefore, an important step in obtaining a fundamental understanding of the origin of instabilities in dynamic fracture. These ideas are as general as the singular behavior that occurs at the tip of a moving crack. It is therefore conceivable that dynamics of the near-tip zone could play an important role in unraveling the physical mechanism driving other instabilities of rapid cracks \cite{Ravi-Chandar.84b,Fineberg.91,Deegan.02.rubber,Livne.07,Scheibert.2010}.

T. G. and J. F. acknowledge the support of the European Research Council (grant 267256) and the Israel Science Foundation (grant 57/07).
E. B. acknowledges support from the Harold Perlman Family Foundation and the
William Z. and Eda Bess Novick Young Scientist Fund. We also thank Mr. Moshe Safran for contributions to the data analysis.

%\bibliography{oscbib}
%\bibliographystyle{apsrev4-1}
%\providecommand{\natexlab}[1]
%\begin{thebibliography}{14}


\onecolumngrid

%\setkeys{Gin}{draft=false}
\renewcommand{\theequation}{A\arabic{equation}}
\vspace{2cm}
\begin{center}
{\large{\bf Supplementary Information}}
\end{center}

\section{Materials}

Our experiments were carried out on cross-linked polyacrylamide gels. Their elastic properties are determined by the molecular weight of the monomer (acrylamide) chains and by the concentration of cross-linking molecules (bis-acrylamide). The gels were prepared by adding to an acrylamide/bis-acrylamide solution (with varying concentrations) two initiators: ammonium persulfate (APS) and tetramethyl ethylene diamine (TEMED). We used a fixed $0.63\%$ (weight/Volume) concentration of the APS while changing the TEMED concentration, thus controlling the length of polymers chains and their respective molecular weight. The following table provides a detailed description of the gels' composition and their measured shear moduli:\\

\begin{tabular}{|c|c|c|c|c|}
\hline & Total Monomer ($\%$) & Cross-Linker/Monomer ($\%$) & Total Initiator (TEMED) ($100 \times \%$) & Shear Modulus, $\mu$ (kPa) \\
\hline 1& 14.1 & 2.7 & 6.2 & $36\pm0.5$ \\
\hline 2& 27 & 4.6 & 6.2 & $168\pm3$ \\
\hline 3& 22.7 & 4.2 & 6.2 & $110\pm6$ \\
\hline 4& 31.4 & 3.2 & 6.2 & $143\pm14$ \\
\hline 5& 14.1 & 2.7 & 3.1 & $34\pm1.5$ \\
\hline 6& 27 & 4.6 & 3.1 & $187\pm17$ \\
\hline
\end{tabular}

\section{Derivation of Eqs. (2)-(3) in the main text}

The polyacrylamide gels in our experiments are incompressible neo-Hookean materials deformed under plane stress conditions. They are described by the following energy functional \cite{S_Knowles.83}
\begin{equation}
U(\B F) = \frac{\mu}{2}\left( F_{ij}F_{ij}+[det(\B F)]^{-2}-3\right) \ ,
\end{equation}
where $\B F$ is the 2D deformation gradient $\B F\!=\! \nabla \B \varphi$ and $\B \varphi$ is the deformation.

As explained in the main text, experimental limitations on the system sizes we could use force us to impose prestrains $\epsilon$ along the y-direction such that the deformation $\B \varphi(\B x)$ takes the form
\begin{eqnarray}
\varphi_x =(1+\epsilon)^{-1/2}x+u_x,\quad\quad\varphi_y =(1+\epsilon)\,y+u_y \ .
\end{eqnarray}
The transverse stretch $\Delta$ is given by
\begin{equation}
\Delta=[det(\B F)]^{-1}=\frac{1}{\pa_x \varphi_x \pa_y \varphi_y-\pa_x \varphi_y \pa_y \varphi_x} \ .
\end{equation}
Note that when $\B u=0$, we have $\Delta=\pa_x\varphi_x=(1+\epsilon)^{-1/2}$, as expected for an isotropic incompressible material under homogeneous uniaxial loading.
The first Piola-Kirchhoff tensor $\B s$, derived from $\B s\!=\!\pa_{\B F} U(\B F)$, reads
\begin{equation}
s_{ij}=\mu\left(\pa_j\varphi_i-\Delta^3\varepsilon_{ik}\varepsilon_{jl}\pa_l\varphi_k \right) \ ,
\end{equation}
where $\varepsilon_{ij}$ is the 2D alternator, i.e. $\varepsilon_{xx}\!=\!\varepsilon_{yy}\!=\!0$, $\varepsilon_{xy}\!=\!-\varepsilon_{yx}\!=\!1$.
The equations of motion $\nabla \cdot \B s \!=\! \rho\,\ddot{\B \varphi}$ can be written explicitly as
\begin{eqnarray}
\label{EOM}
\mu\nabla^2\varphi_x+\mu\left[\pa_y(\Delta^3)\pa_x \varphi_y-\pa_x(\Delta^3)\pa_y \varphi_y \right]&=&\rho\,\ddot{\varphi_x}\ ,\\
\label{EOM1}
\mu\nabla^2\varphi_y+\mu\left[\pa_x(\Delta^3)\pa_y \varphi_x-\pa_y(\Delta^3)\pa_x \varphi_x \right]&=&\rho\,\ddot{\varphi_y} \ ,
\end{eqnarray}
where $\rho$ is the reference mass density.
The traction-free boundary conditions on the crack faces $s_{xy}(r,\theta\!=\!\pm\pi)\!=\!s_{yy}(r,\theta\!=\!\pm\pi)\!=\!0$
can be written explicitly as
\begin{equation}
\label{BC}
s_{xy}(r,\theta=\pm\pi)=\mu(\pa_y\varphi_x+\Delta^3\pa_x\varphi_y)|_{\theta=\pm\pi}=0,\quad\quad
s_{yy}(r,\theta=\pm\pi)=\mu(\pa_y\varphi_y-\Delta^3\pa_x\varphi_x)|_{\theta=\pm\pi}=0 \ ,
\end{equation}
where the polar coordinate system $(r,\theta)$ is moving with the crack tip such that $\theta\!=\!0$ is the crack tip propagation direction.

Focusing on steady state conditions, $\pa_t\!=\!-v\pa_x$, we can now linearize Eqs. (\ref{EOM}), (\ref{EOM1}) and (\ref{BC}) with respect to $\B u$.
The results read
\begin{eqnarray}
\label{1st_x}
4\pa_{xx} u_x + 3(1+\epsilon)^{-3/2} \pa_{xy} u_y + \pa_{yy} u_x - v^2 \pa_{xx} u_x &=& 0 \ ,\\
\label{1st_y}
\pa_{xx}u_y + 3(1+\epsilon)^{-3/2} \pa_{xy} u_x + \left[ 1+ 3\left( 1+\epsilon\right) ^{-3}\right]  \pa_{yy} u_y - v^2 \pa_{xx} u_y &=& 0\ ,
\end{eqnarray}
with the following boundary conditions at $\theta=\pm\pi$
\begin{eqnarray}
\label{1st_x_BC}
\pa_y u_x + (1+\epsilon)^{-3/2} \pa_x u_y&=& 0\ ,\\
\label{1st_y_BC}
\left[1+ 3\left( 1+\epsilon\right)^{-3}\right]  \pa_{y} u_y + 2(1+\epsilon)^{-3/2} \pa_x u_x &=& 0 \ ,
\end{eqnarray}
where $v$ here is normalized by $c_s=\sqrt{\mu/\rho}$. In the limit $\epsilon \to 0$, we recover the equations of isotropic LEFM.
The two-term asymptotic expansion near the crack tip appears in Eq. (2) in the main text and is copied here
\begin{eqnarray}
\hspace{-0.5cm}&&u_x(r,\theta; v,\epsilon)=\frac{K_I(v,\epsilon) \sqrt{r}}{4\mu\sqrt{2\pi}}\Omega_x(\theta;v,\epsilon)\!+\!\frac{T_x(\epsilon)\,r\cos\theta}{\mu},\nonumber\\
\label{firstO}
\hspace{-0.5cm}&&u_y(r,\theta; v,\epsilon)=\frac{K_I(v,\epsilon)\sqrt{r}}{4\mu\sqrt{2\pi}}\Omega_y(\theta;v,\epsilon)\!+\!\frac{T_y(\epsilon)\,r\sin\theta}{\mu} \ ,
\end{eqnarray}
The function $\B \Omega(\theta;v,\epsilon)$ is calculated using the following half-integer Fourier expansion
\begin{eqnarray}
\label{pre_sqrt_r}
\Omega_x(\theta;v,\epsilon) = \sum_n a_n(v,\epsilon) \cos\left[\frac{(2n-1)\,\theta}{2} \right], \quad\quad\Omega_y(\theta;v,\epsilon) = \sum_n b_n(v,\epsilon) \sin\left[\frac{(2n-1)\,\theta}{2} \right]\ .
\end{eqnarray}
$\{a_n,b_n\}$ are calculated by solving a set of linear algebraic equations for each $v$ and $\epsilon$, where $n$ is selected such that the required accuracy is obtained. Note that since this is a linear set of equations, $\{a_n,b_n\}$ can be determined up to an overall multiplicative factor, which as usual is quantified by the stress intensity factor $K_I$ that cannot be calculated by the asymptotic analysis (but rather from the global crack problem). The standard LEFM result is recovered for $\epsilon\to 0$. For the subleading term we obtain
\begin{eqnarray}
\label{pre_r}
T_x(\epsilon) = \frac{T(\epsilon)[1+3\left( 1+\epsilon\right)^{-3}]}{12}, \quad T_y(\epsilon) = -\frac{T(\epsilon)\left( 1+\epsilon\right)^{-3/2}}{6} \ ,
\end{eqnarray}
where $T(\epsilon)$ cannot be determined by the asymptotic analysis. Note that, again, the standard LEFM result is recovered for $\epsilon\to 0$.
In the absence of a global nonlinear solution of our problem, we use the latter to obtain a sensible estimate of $T_x(\epsilon)$, which appears in Eq. (3) in the main text. It is known that for a tensile crack in a large system under remote tensile stress $\sigma^\infty$, $T_x(\epsilon)/\mu\!=\!-\sigma^\infty/E$, where $E$ is the Young's modulus \cite{S_99Bro}. Since the background strains in our experiments were of the order of $0.1$, we use the linear approximation $\sigma^\infty\!\simeq\!\epsilon E$ to obtain $T_x(\epsilon)/\mu\!\simeq\!-\epsilon$. We used this estimate in our analysis.

To obtain the crack tip opening displacement (CTOD) we focus on $\theta\!=\!\pi$, which leads to
\begin{eqnarray}
\varphi_x(r,\pi)=-r\left[(1+\epsilon)^{-1/2}-\epsilon\right],\quad\varphi_y(r,\pi) =\frac{K_I\sqrt{r}}{4\mu\sqrt{2\pi}}\Omega_y(\pi;v,\epsilon)\ .
\end{eqnarray}
Eliminating $r$ between the last two relations, we obtain a parabolic form $\varphi_x(r,\pi)\!=\!-a\,\varphi_y^2(r,\pi)$ with
\begin{equation}
\label{curvature}
a=\left[(1+\epsilon)^{-1/2}-\epsilon\right]\dfrac{32\pi\mu^2}{\Omega_y^2(\pi;v,\epsilon)K_I^2}\ .
\end{equation}
The final step in deriving Eq. (3) in the main text would be to consider the J-integral \cite{S_98Fre}
\begin{equation}
\label{Gint}
G=\int_C \left[\left(U+\frac{1}{2}\rho v^2\pa_x u_i \pa_x u_i\right) n_x- s_{ij}n_j\pa_x u_i\right]dC\ ,
\end{equation}
where $C$ is any path surrounding the crack tip. We calculated the integral numerically using Eqs. (\ref{firstO}), which allows us to define $A(v;\epsilon)$ from the relation
\begin{equation}
G = \frac{A(v;\epsilon) K_I^2(v,\epsilon)}{32\pi \mu^2} \ .
\end{equation}
Using energy balance, $\Gamma(v)\!=\!G$, together with Eq. (\ref{curvature}), we arrive at Eq. (3) in the main text.

\section{Additional comments}

\begin{itemize}
\item {\bf The constant of proportionality between $\lambda_{osc}$ and $\delta(v_c)$}\\
\vspace{0.3cm}
The constant of proportionality between $\lambda_{osc}$ and $\delta(v_c)$ in Fig. 4 in the main text is about $10$. To understand this, consider Eq. (14) in \cite{S_09Bou}
\begin{equation}
\label{ini_wave}
\lambda_{osc} = \frac{2 \pi v_c \beta}{c_{nl}\Re{(\bar{\omega}_c)}}\ell_{nl}(v_c) \ ,
\end{equation}
where $c_{nl}$ is the typical wave speed within the nonlinear zone and $\beta$ is a dimensionless number quantifying the typical time it takes mechanical information to propagate across the nonlinear zone (and is order unity). The real part of the dimensionless complex oscillation frequency was estimated to be $\Re{(\bar{\omega}_c)}\!\simeq\!1.5$. Furthermore, $\beta\!\simeq\!1/2$ and $c_{nl}\!\simeq\!c_s$ was used, leading to $\lambda_{osc}\!\simeq\!1.6\,\ell_{nl}(v_c)$ \cite{S_09Bou}. Finally, $\ell_{nl}(v_c)$ was estimated using a weakly nonlinear calculation {\em ahead} of the crack tip \cite{S_08BLF} to be of the order of a few mm. The relation $\lambda_{osc}\!\simeq\!10\,\delta(v_c)$ may emerge from two observations: (i) $\delta(v_c)$ is measured {\em behind} the crack tip, that may be somewhat smaller than the estimate {\em ahead} of the tip, leading to $\ell_{nl}(v_c)\!=\!\alpha\,\delta(v_c)$, with $\alpha\!>\!1$ (ii) $\beta$ may be more faithfully estimated as $\beta\!=\!2$, which corresponds to a round trip across the nonlinear zone, increasing the prefactor $1.6$ in $\lambda_{osc}\!\simeq\!1.6\,\ell_{nl}(v_c)$ to about $6$.

\item{\bf The nonlinear scale $\Gamma/\mu$}\\
\vspace{0.3cm}
In the main text the weakly nonlinear scale $\Gamma/\mu$ is used, e.g. Fig. 2. We would like to stress that in the most general case a weakly nonlinear elastic lengthscale depends on a set of second order elastic coefficients, in addition to the linear elastic coefficient $\mu$ and $\lambda$ (the first Lam\'e constant). To see this, write the following expansion of the energy functional $U$ in 2D
\begin{eqnarray}
\label{U_E_2d}
\hspace{-0.4cm}U^{2D}(\B E) = \frac{\lambda}{2} (tr\!\B E)^2 + \mu\,tr\!\B E^2 + \frac{l}{3} (tr\!\B E)^3 +\frac{2m}{3}tr\!\B E^3  \ ,
\end{eqnarray}
where the metric (Green-Lagrange) strain measure $\B E \!=\! \case{1}{2}(\B F^T \B F \!-\!\B I)$ is understood as a 2D tensor and $\{l,m\}$ are two of the Murnaghan coefficients \cite{S_51Murnaghan}. In 3D there is an additional second order coefficient $n$. Eq. (\ref{U_E_2d}) reduces to Eq. (1) in the main text when $\lambda=2\mu,\;l=-4\mu,\;m=-4\mu$. For other materials $l$ and $m$ may be larger. For example, for Polystyrene (a glassy polymer) $l \!\simeq\! -14 \mu$ \cite{S_53HK}. Since the weakly nonlinear estimate for $\ell_{nl}$ involves $(l/\mu)^2$ \cite{S_BFM10}, $\Gamma/\mu$ (which is of course dimensionally correct) may include widely varying prefactors for different materials.

\end{itemize}

\end{document}